# Fractal-Based Electrolytic Capacitor Electrodes: Scaling Behavior with Respect to Fractal Order and Complexity

Benjamin Barnes[1,2], Othman Suleiman[1], JeanPaul Badjo[1] and Kausik S Das[1]

[1]University of Maryland Eastern Shore,
1, Backbone Road, Princess Anne, MD 21853 USA
[2]Department of Chemistry and Biochemistry,
University of Maryland College Park, MD 20742 USA

In past decades, the application of fractals to electrode design for enhanced signaling and electrochemical performance was a popular subject and enabled the growth of consumer micro-electronics. Supercapacitors, which are energy storage devices with many promising characteristics, have largely grown alongside of such developments in electronics, but little work has been done to use fractal electrodes in supercapacitors. In this work, plane-filling and fractal patterns were used in designing laser scribed graphene supercapacitor electrodes, allowing the scaling laws of capacitance with respect to fractal order and complexity to be examined for the first time. An interesting exponential relationship between capacitance and fractal order for the more open structured fractals was observed, the exponent of which was proportional to the Hausdorff dimension. Additional non-linear relationships between capacitance and order were observed for other structures which was correlated with inter-plate repulsion and differences in path length. These findings provide the first step in maximizing the efficiency of fractal-based electrolytic devices by exploring the non-intuitive trends in capacitance with respect to fractal order and complexity.

Due to their rapid charge/discharge rate, long operational lifetime[1], and potentially flexible and environmentally benign designs[2], electrolytic supercapacitors (ESCs) are poised to become an integral component of a range of electronic systems ranging from the acceleration boosters in electric cars to flexible cellular phones[3]. The primary focus of recent ESC research has been in developing new active materials, which include activated carbon,[1] carbon nanotubes [4,5], carbon nano onions[6], and carbide-derived carbons[6]. Graphene, however, received the greatest attention[6,7,8,9,10] due to its high surface area and conductivity and chemical stability [9]. A number of recent papers have propelled the method of fabricating ESCs from laser-reduced graphene oxide (GO) films into the realm of commercial feasibility[11,12]. This method is attractive as it serves to both exfoliate and reduce the GO resulting in the liberation of oxygen as $CO$, $CO_2$, and $H_2O$ gas[11,12]. By tuning such parameters as laser intensity, scribing speed, and substrate temperature, the conductivity and overall quality of the laser scribed graphene (LSG) layer can be precisely controlled[12,13,14,15], but by far, the greatest benefit of this technology is the unprecedented control over the detail of structures fabricated thanks to the high precision of modern laser scribing instruments[16,17]. This precision has not been fully exploited though, and most researchers continue to produce simple interdigitated, spiral, concentric, or parallel-bar ESC electrodes in their prototypes[18,**Error! Bookmark not defined.**,19,19]. The benefit of using more complex electrode designs, namely fractal patterns, remains largely a virgin field, with no work done to determine the relationship between fractal order and electrolytic capacitance. This is despite the success of applying fractal patterns to other fields in electronics[20,21] and the strong theoretical case for fractal ESCs[22]

In this work, we demonstrate the impact of using intricate LSG fractal patterns in ESC electrodes. We examine the scaling laws which govern the relationship between fractal order and capacitive performance for several different fractal and plane-filling curves.

**Results and Discussion**

Due to external pressures such as natural selection, nature is expert in fully exploiting available surfaces by the formation of intricate designs, for example by the repetition of similar patterns over a great range of length scales: such structure is described by fractal geometry. As a result, fractals, and related space-filling structures, can be found throughout nature in the patterns of clouds[23], neuronal and vessel networks[24], as well as coasts, trees, galaxy clusters, leaves, and the paths of particles undergoing Brownian motion, to name but a few of the diverse topics investigated and popularized by Mandelbrot[25]. Fractals are defined as objects possessing a Hausdorff dimension greater than that of their geometric dimension. The Hausdorff dimension of a fractal is calculated based on scalability and self-similarity using the following formula:

$$Hausdorff\ Dimension = \frac{\log(R)}{\log(M)}$$

Where R is the repetition of self-similar patterns for each iteration of the fractal order, and M is the factor by which the self-similar pattern decreases in size for each iteration. The Hausdorff dimension, therefore, can be viewed as a rough measure of the compactness and complexity of the particular fractal structure under investigation. The Hausdorff dimension is shown in Figure 1a for each fractal investigated in this work. It is also shown in Figure 1b that the Hausdorff dimension is directly proportional to the path-length function of the fractal pattern.

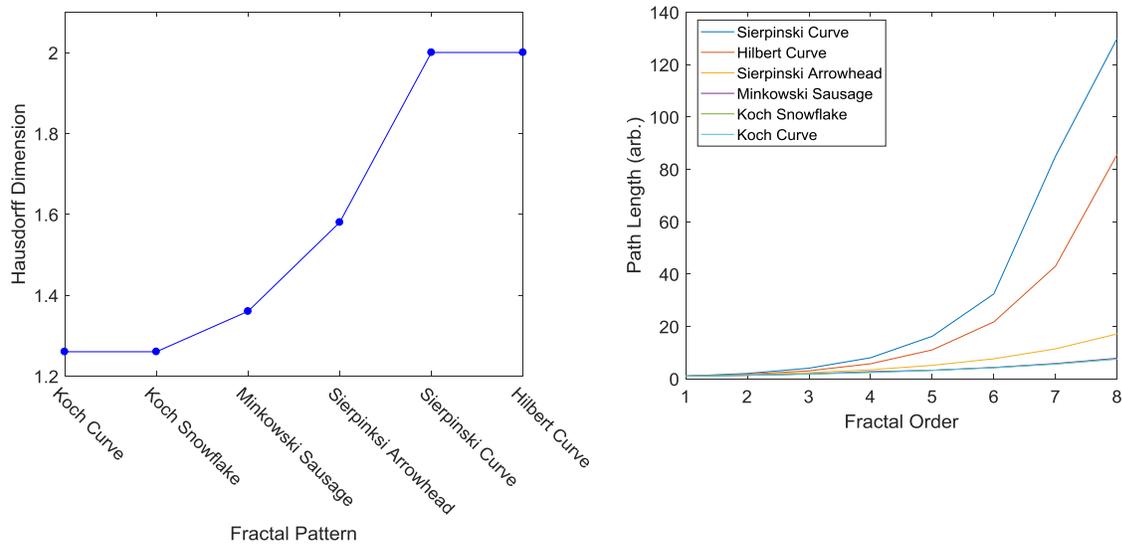

*Figure 1: a) Hausdorff dimension for a number of fractal and plane-filling curves; b) relationship between path length and fractal order for the same structures*

From Figure 1 it can be concluded that path length increases exponentially with order for all of the fractal structures investigated in this work, but does so more significantly for those structures with a larger Hausdorff dimension, namely the Sierpinski and Hilbert curves. Area should follow the same trend as it is simply the product of path length and the constant width of the line used to draw the fractal. Therefore, since electrolytic capacitance is related to area through the following formula:

$$C = \frac{kS}{d}$$

where C is capacitance, k is the dielectric constant for the electrolyte, S is surface area, and d is the electrolytic double layer (EDL) thickness, it would be expected that capacitance would also increase exponentially for capacitors made with fractal electrodes. This would be true in an ideal case, but in the real world, other factors are involved which are detrimental to capacitance, such as path resistance, and the effect of coulombic repulsion between adjacent portions of the like-plates of the fractal pattern. These trends were investigated by fabricating electrolytic capacitors of different orders for the fractal and space-filling curves in Figure 1.

The fractals chosen for this work are popular and representative structures in mathematics, and are shown below in Figure 2. These are the Hilbert curve, Koch curve, Sierpinski curve, Minkowski sausage, Koch snowflake, and the Sierpinski arrowhead.

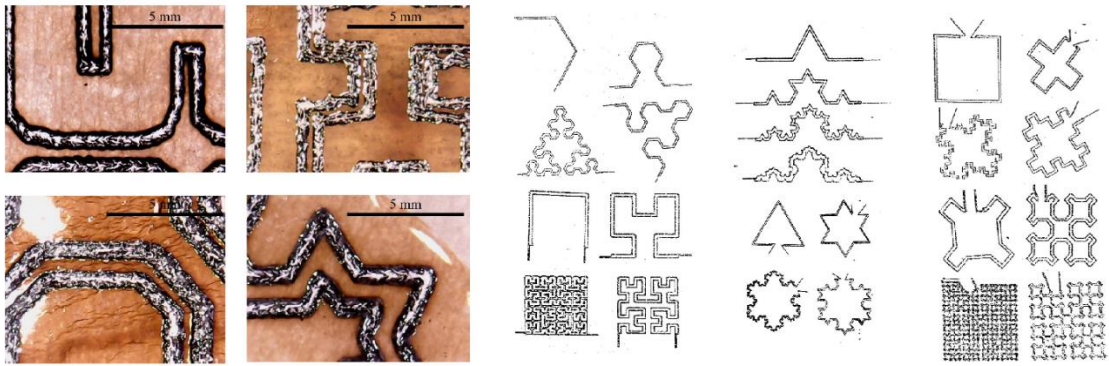

*Figure 2: Fractal electrode designs and detail of the laser scribed fractal pattern. (L-R): detail of scribed pattern, Sierpinski arrowhead and Hilbert curves, Koch and Koch snowflake curves, Minkowski and Sierpinski curves.*

The fractal patterns were scribed in a ~30 µm film of graphene oxide which had been drop-casted on a PET substrate. The scribing was accomplished using a 450 nm laser scribing system. The reduction of GO to LSG is evidenced by a dramatic reduction in resistance from 12 000 Ω/mm² for the GO sheet to 670 Ω/mm for the laser reduced graphene pattern. The scribing system is capable of great detail and continuity as seen in the inset above, making it an ideal means of fabricating fractal electrodes. The ESC was completed by depositing a film of solid state electrolyte over its surface. This process is summarized in Figure 3.

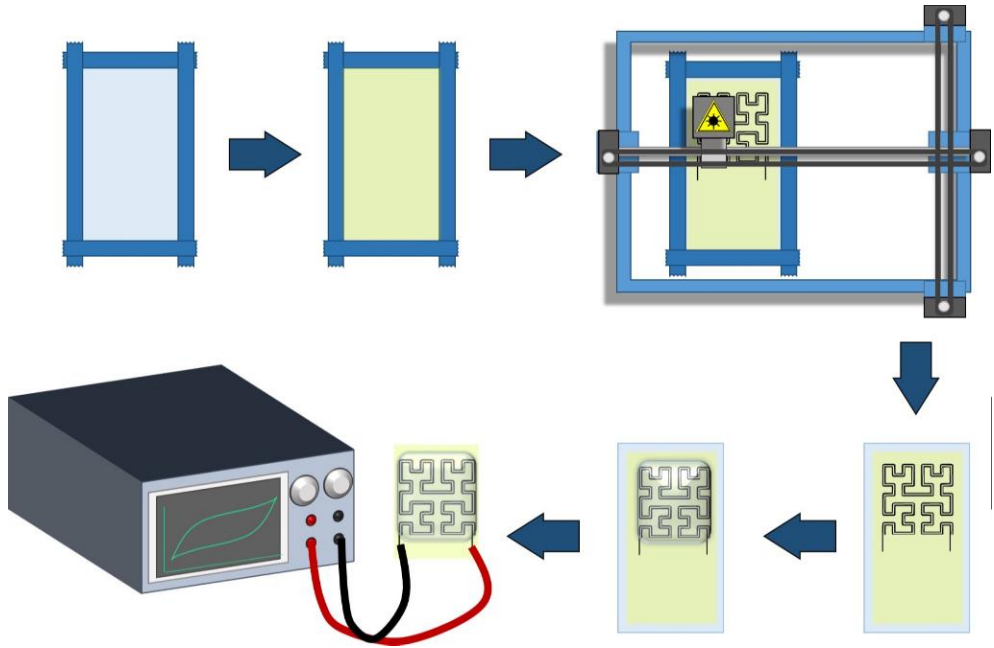

*Figure 3: Method of fabricating fractal electrodes for electrolytic supercapacitors. (Clockwise from top left): A PET film was cleaned with isopropanol and masked; GO solution was deposited on the PET film and allowed to dry; the desired fractal pattern was scribed into the GO film using a x-y plotting system; the electrode sheet was removed from the x-y plotter; a film of solid-state PVA-phosphoric acid electrolyte was deposited on the electrode; the electrode was evaluated using an LCR meter and by analyzing its CV behavior.*

FTIR and Raman spectroscopy also showed the expected changes indicative of laser reduction of GO to RGO, namely a loss of the –OH band at 3150 cm$^{-1}$ and the -O- band at 1045 cm$^{-1}$ in the FTIR and a shift in the relative heights of the D and G bands in the Raman spectrum. Both spectra are shown below in Figure 4.

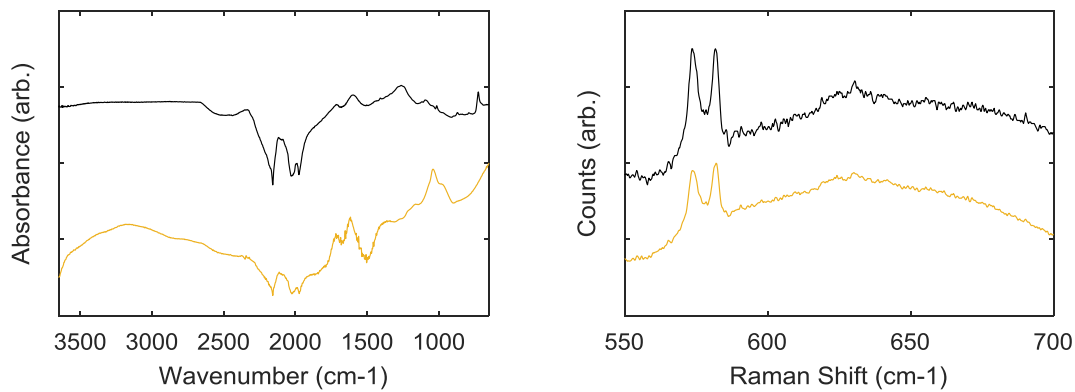

*Figure 4: (L-R): FTIR and Raman spectra showing evidence of GO reduction to conducting RGO.*

The first ESCs to be evaluated were the Sierpinski curve and Minkowski sausage. These fractals have a Hausdorff Dimension of 2.00 and 1.50, respectively, indicating that these two structures scale with similar complexity. It is also notable that for lower fractal orders (1st, and 2nd), the two structures are visually similar, with a large open space in the centers of the first order structure, which is divided into roughly a cross shape in the second order (Figure 5). The similarity of the two patterns then diverges for higher orders, with the Sierpinski curve becoming more complex overall and the Minkowski Sausage remaining roughly cross- or sauwastika-shaped with a more open center. These changes can be easily visualized in Figure 5.

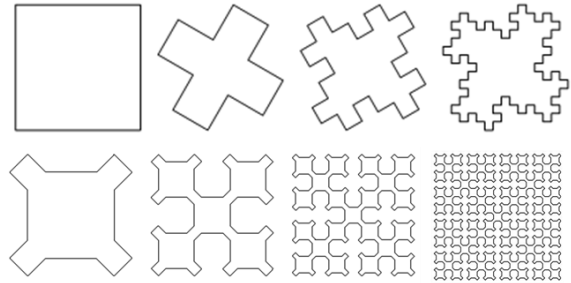

Figure 5: Relative geometry of the four orders of the Minkowski curve (top) and the Sierpinski curve (bottom)

This drastic increase in the complexity of the Sierpinski curve is due to its space-filling nature, meaning that as the order tends to infinity, the curve tends to cross every point in the x-y plane. Figure 6 shows the respective relationships between capacitance and fractal order for the Sierpinski curve (top) and the Minkowski curve (bottom).

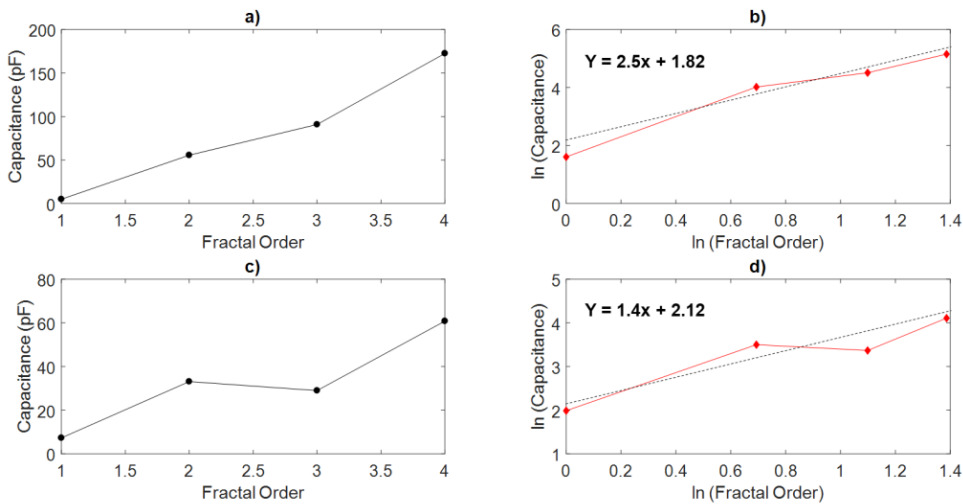

Figure 6: Changes in capacitance with respect to changes in fractal order for a) the Sierpinski curve and c) the Minkowski curve. A log-log analysis of the exponential growth of capacitance of both curves in seen in b) and c) respectively.

The exponential nature of the capacitance-order relationship for both of these fractals is seen in plots a) and c), and is further investigated by plotting the respective natural logs of the ordinate and abscissa shown in plots b) and d). Such near-exponential relationships are conveniently described by a power law such as follows:

$$y = ax^k$$

Where, in this case, y is capacitance, x is fractal order, k is the scaling factor, and a is a constant. The slope of the log-log plot corresponds to k, while the x intercept is taken to be ln(a). From this information, a scaling law can be formed which describes the relationship between capacitance (y) and fractal order (x) for the Sierpinski curve,

$$y = 6.11x^{2.52}$$

And the Minkowski curve.

$$y = 8.33x^{1.40}$$

It is notable that the power in the scaling law in this case is proportional to the Hausdorff dimension, with the Sierpinski curve having a larger exponent and a larger Hausdorff dimension. It is tempting to assume that this relationship holds true in general, but there are likely many other factors involved in determining capacitance, such as repulsion due to the proximity of like-charged plates, and the effect of increasing resistance with increasing path length.

To determine how these two factors influence the capacitance-order relationship, two additional fractal structures were fabricated: the Koch curve and the Koch snowflake. The benefit of using these two structures is that he Hausdroff dimension and dimensional scaling law of these two are identical, and the path length differs by a constant factor of 3. The only change with increasing order then, is in the magnitude of repulsions between like plates; the Koch curve remains a highly open structure with minimal repulsion despite changes in order, whereas the Koch snowflake is cyclic, so additional increases in order result in increasingly close contact between any two points on the like plates of the capacitor and therefore increased coulombic repulsion.

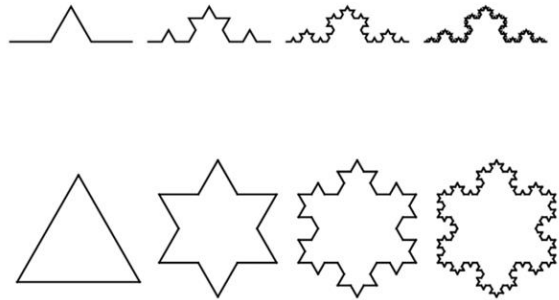

Figure 7: Comparison of the geometry of the first four iterations of the Koch curve (top) and the Koch snowflake (bottom). The Koch curve is clearly a more open structure with little repulsion, whereas the Koch snowflake's cyclic structure is more repulsive.

This repulsion would be expected to manifest itself in increased capacitance for changes in order for the Koch curve, but decreased capacitance for changes in order for the Koch snowflake.

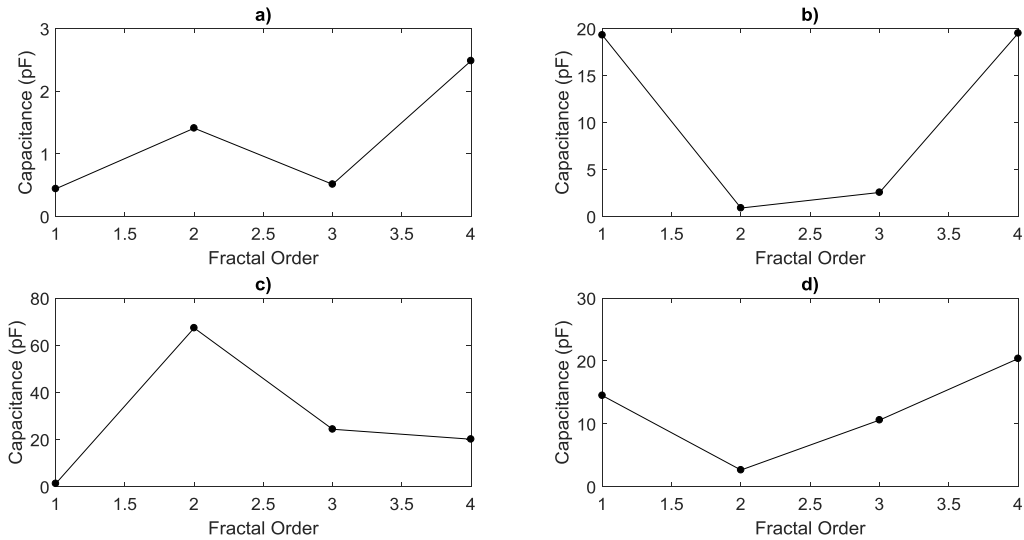

*Figure 8: Comparison of the capacitance-order relationship for a) the Koch curve c) the Koch snowflake, b) the Sierpinski arrowhead, and d) the Hilbert curve*

As shown in Figure 8a, and c above, the predicted trend was generally observed, with capacitance increasing in a more linear fashion for the Koch curve (a) with each change in order. The results were much more variable for the Koch snowflake (b), as expected; there was a promising initial increase in capacitance between order 1 and 2 which then declined and plateaued for orders 3 and 4. This indicates that initially there was a similar relationship between capacitance and order as for the Koch curve, but as order increased, the repulsions between adjacent portions of the like plates overcame the growth in capacitance for the Koch snowflake.

Next the effect of surface area was investigated in two intermediate fractal patterns: these were the Hilbert curve and the Sierpinski arrowhead. Both curves possess large Hausdorff dimensions (2.00 and 1.58, respectively), and are therefore relatively compact and complex structures. The primary difference between the two structures is the rate of change in path length, and therefore surface area, with increasing order. As seen in Figure 1b, Hilbert curve has a greater exponential growth in length than does the Sierpinski arrowhead.

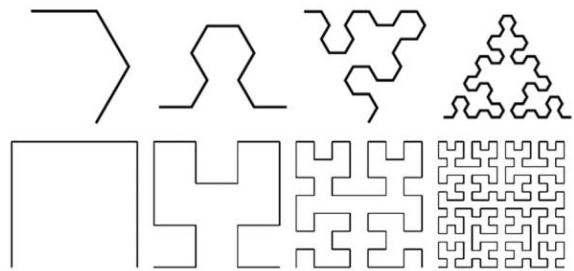

*Figure 9: Comparison of the geometries of the first four orders of the Sierpinski arrowhead (top) and the Hilbert curve (bottom)*

The capacitance and order data Figure 8 (b and d) are again non-linear, and *prima facie,* non-intuitive. In both cases, capacitance begins at a relatively large value for the first order structures. This is expected as both structures possess large surface area and in the first order, there is little repulsion due to the great distance between like plates. However, when the order is increased to 2, there is a large decline in capacitance for both fractals due, presumably, to the increased repulsions as anticipated. However, the more interesting result for this set is the increase in capacitance following the initial slump. This is especially evident for the Hilbert curve, capacitance of which begins increasing again after the $2^{nd}$ order. This is likely due to the large exponential increase in surface area with order Figure 1b which may be capable of overcoming the repulsions due the compact structure. This exponential increase in surface area with order is not as large for the Sierpinski arrowhead, and therefore there is a delay in the recovery of capacitance which doesn't begin until the $3^{rd}$ order.

Taken together, these results illustrate an interesting circumstance: for some simple fractal structures with high surface area such as the Sierpinski curve, or with large open structures as in the Minkowski sausage, there exists a linear or exponential relationship between electrochemical capacitance and fractal order. This relationship is shown to be roughly proportional to the Hausdorff dimension of the particular fractal under investigation. As the structure becomes more complex, though, other forces seem to limit the increase in capacitance. This was shown by comparing the Koch curve and Koch snowflake, which share the same Hausdorff dimension and length scaling law, but differ in the magnitude of interaction between like plates. Thus the capacitance scales differently between these two structures. Additionally, by comparing the capacitance scaling of the Hilbert curve with the Sierpinski arrowhead, it was shown that in some cases at least, the repulsive effects can be overcome at higher orders by a structure which increases its surface area at a sufficient rate.

**Methods**

GO films were prepared on thin PET substrates by drop casting 6.7 mL of aqueous GO (Graphenea, 4 g/L) on a ~14 400 $mm^2$ sheet (cleaned with isopropyl alcohol). The GO film and PET substrate was incubated at 50°C until dry, cut into sheets, and stored. The areal density of GO on the substrate was calculated to be ~$2.7\times10^{-6}$ g $mm^{-2}$. The average thickness of these films was ~30 µm. The GO films appeared to be dark gray to ochre, depending on the thickness, and had a transmittance of ~45% of the solar spectrum. The GO film had a sheet resistance of ~12 000 Ω/$mm^2$. The adhesion between the GO film and the PET substrate was poor, but adhesion between the LSG and PET was much better, possibly due to GO being fused into the partially molten plastic during the laser scribing process. Resistance in the LSG path was ~670 Ω/mm. FTIR and Raman analysis of the GO film revealed the strong OH stretching band centered at 3150, G and D bands at 1720 and 1615, respectively, and an ether peak at 1045. These bands and peaks were largely flattened when the GO film was laser scribed, indicating a successful reduction of the film as described in previous literature[26]; further evidence of the reduction process can be seen in the Raman spectrum as a shift in the relative heights of the D and G bands.

Sierpinski, Minkowski, Koch, Koch snowflake, Hilbert and Sierpinski arrowhead curves of the $1^{st}$ through $4^{th}$ order were plotted using Wolfram Mathematica's built-in functions of the same name. The resulting graphics items was copied and opened in a drawing software (Microsoft Paint) where simple current collectors were added to the image. The electrode image was then opened in BenBox laser scribing software which was linked to an x-y plotter (MakeBlock x-y plotter V 2.0). A laser setting of 200 units of intensity, and a scribing speed of 6.7 mm/s was used. The laser used was supplied by MakeBlock and was a 450 nm blue-violet laser with a maximum power output of 450 mW – 500 mW. By using the "scan by outline" setting in BenBox, the inner and outer outline of the fractal image was scribed, resulting in a double fractal pattern which is parallel at each point. The inner and outer lines could then be attached to positive and negative current collectors to complete the capacitive circuit. Theoretically this design would enable both electrolytic capacitance between the LSG and electrolyte in addition to traditional capacitance by the storage of opposite charges on parallel LSG regions. The distance between the parallel lines was easily controlled by the Thickness function in Mathematica. The GO layer was scribed 4 times, as previous work has shown that multiple laser passes improve conductivity[14].

A polymer gel electrolyte (PVA-phosphoric acid) was prepared as described elsewhere[27] by combining PVA powder, water, and concentrated phosphoric acid in a 10:100:8 mass ratio. The resulting thick solution was pipetted across the scribed electrode (excluding only the current collector region), which was then stored away from dust until the gel was dry, a process which took up to 24 hours depending on temperature, humidity, and airflow. The final resistance of the electrolyte (measured by 4-point probe) was ~80 $\Omega/mm^2$ and the thickness of the layer (measured by Keyence LK Laser Displacement Meter) was approximately 35 µm. The average volume of these devices was approximately 0.32 $cm^3$ (measured by water displacement) and mass was about 0.32 g.

The completed ESC was evaluated using a BK Precision 891 LCR Meter, and a Keithley 2450 SourceMeter. Capacitance (C), Q factor (Q), equivalent series resistance (R), and dissipation factor (D) were determined at 200 Hz, 1 VRMS, in FAST mode using the LCR meter. The cyclic voltammetry (CV) behaviour of the capacitors was determined with the SourceMeter at different sweep frequencies between 0 and 1 V and 10 or 50 double sweeps.

**Acknowledgement**

KD and J.P. Badjo would like to thank Maryland Technology Enterprise Institute for their generous support through MIPS grant.